\documentclass[a4paper]{article}
\usepackage{enumitem,hyperref,dahlin,natbib,multirow,rotating,sectsty,booktabs}              
\allsectionsfont{\sffamily}
\usepackage[margin=1in]{geometry}
\usepackage{acronym}

\newcommand{\Transp}{\mathsf{T}}
\DeclareMathOperator{\Tr}{Tr}

\newcommand{\gth}{g_\theta}
\newcommand{\fth}{f_\theta}

\acrodef{map}[MAP]{maximum a posteriori}
\acrodef{ml}[ML]{maximum likelihood}
\acrodef{ekf}[EKF]{extended Kalman filter}
\acrodef{eks}[EKS]{extended Kalman smoother}
\acrodef{kf}[KF]{Kalman filter}
\acrodef{bfgs}[BFGS]{Broyden-Fletcher-Goldfarb-Shanno}
\acrodef{rts}[RTS]{Rauch-Tung-Striebel}
\acrodef{pdf}[PDF]{probability density function}
\acrodef{ssm}[SSM]{state space model}
\acrodef{FL}[FL]{fixed-lag}
\acrodef{pf}[PF]{particle filter}
\acrodef{FFBSi}[FFBSi]{forward filter backward simulator}
\acrodef{SSm}[ssm]{state space model}
\acrodef{EM}[EM]{expectation-maximization}

\newlength\myindent
\usepackage{setspace}

\title{Newton-based maximum likelihood estimation \\ in nonlinear state space models} 

\author{Manon Kok%
\thanks{Department of Electrical Engineering, Link{\"o}ping University, Link{\"o}ping, Sweden. E-mail: \textit{manon.kok@liu.se}.} 
, Johan Dahlin%
\thanks{Department of Electrical Engineering, Link{\"o}ping University, Link{\"o}ping, Sweden. E-mail: \textit{johan.dahlin@liu.se}.} 
, Thomas B.\ Sch\"{o}n%
\thanks{Department of Information Technology, Uppsala University, Uppsala, Sweden. E-mail: \textit{thomas.schon@it.uu.se}.}
\, and Adrian Wills%
\thanks{School of Engineering, University of Newcastle, Australia. E-mail: \textit{adrian.g.wills@gmail.com}.%
}}

\begin{document}

\maketitle

\doublespacing
\begin{abstract}
\noindent Maximum likelihood (ML) estimation using Newton's method in nonlinear state space models
  (SSMs) is a challenging problem due to the analytical intractability of the log-likelihood and its gradient and Hessian. We estimate the gradient and Hessian
  using Fisher's identity in combination with a smoothing algorithm. We explore two approximations of the log-likelihood and of the solution of the smoothing problem. The first is a linearization approximation which is computationally cheap, but the accuracy typically varies between models. The second is a sampling approximation which is asymptotically valid for any SSM but is more computationally costly. We demonstrate our approach for ML parameter estimation on simulated data from two different SSMs with encouraging results.\\

\noindent \textbf{Keywords}: Maximum likelihood, parameter estimation, nonlinear state space models, Fisher's identity, extended Kalman filters, particle methods, Newton optimization. \\ \\

\noindent The data and source code are available at \url{http://users.isy.liu.se/en/rt/manko/} and GitHub: \url{https://github.com/compops/newton-sysid2015}.
\end{abstract}

%%%%%%%%%%%%%%%%%%%%%%%%%%%%%%%%%%%%%%%%%%%%%%%%%%%%%%%%%%%%%%%%%%%%%%%%%%%%%%%%%%%%%%%%%%%%%%%%%%%%%%%
%%%%%%%%%%%%%%%%%%%%%%%%%%%%%%%%%%%%%%%%%%%%%%%%%%%%%%%%%%%%%%%%%%%%%%%%%%%%%%%%%%%%%%%%%%%%%%%%%%%%%%%
%%%%%%%%%%%%%%%%%%%%%%%%%%%%%%%%%%%%%%%%%%%%%%%%%%%%%%%%%%%%%%%%%%%%%%%%%%%%%%%%%%%%%%%%%%%%%%%%%%%%%%%
\newpage
\section{Introduction}
% Introduce the state space model and the corresponding ML parameter inference problem
Maximum likelihood (ML) parameter estimation is a ubiquitous problem in control and system identification, see e.g.\ \cite{ljung:1999} for an overview. We focus on ML estimation in nonlinear state space models (SSMs), \acused{ssm}\acused{ml}
\begin{equation}
\begin{split}
\label{eq:ss}
x_{t+1} &= \fth  (x_t, v_t), \\
y_t &= \gth (x_t, e_t), 
\end{split}
\end{equation}
where $x_t$ and $y_t$ denote the \textit{latent} state variable and the measurement at time $t$,
respectively. The functions $\fth ( \cdot )$ and $\gth ( \cdot )$ denote the dynamic and the measurement
model, respectively, including the noise terms denoted by $v_t$ and $e_t$. The ML estimate of the static parameter vector $\theta \in \Theta \subseteq \mathbb{R}^p$ is obtained by solving
\begin{align}
\label{eq:ml}
\widehat{ \theta}_\text{ML} &= \argmax_\theta \ell_\theta (y_{1:N}),
\end{align}
where $\ell_\theta (y_{1:N}) \triangleq \log p_\theta ( y_{1:N} )$ denotes the log-likelihood function and $y_{1:N}=\{y_1,\ldots,y_N\}$.

In this paper, we aim to solve \eqref{eq:ml} using Newton methods~\citep{nocedalW:2006}. These enjoy quadratic convergence rates but require
estimates of the log-likelihood and its gradient and Hessian. For linear Gaussian SSMs, we can
compute these quantities exactly by the use of a Kalman filter (KF; \citealp{kalman:1960}) and by using a Kalman
smoother together with \textit{Fisher's identity} \citep{fisher:1925}. This approach has previously
been explored by e.g.\ \cite{segalW:1989}. An alternative approach is to compute the gradient
recursively via the sensitivity derivatives \citep{astrom:1980}. However, neither of these
approaches can be applied for general SSMs as they require us to solve the analytically intractable
state estimation problem for~\eqref{eq:ss}. \acused{kf}

The main contribution of this paper is an extension of the results by \cite{segalW:1989} to general SSMs in order to solve \eqref{eq:ml}. To this end, we explore two approximations of the log-likelihood and the solution of the smoothing problem, using linearization and using sampling methods. The \textit{linearization} approximation estimates the log-likelihood using an extended Kalman filter (EKF), see e.g.\ \citet{gustafsson:2012}. The smoothing problem is solved using Gauss-Newton optimization. The \textit{sampling} approximation is based on particle filtering and smoothing \citep{DoucetJohansen2011}. \acused{ekf}

The main differences between these two approximations are the accuracy of the estimates and the computational complexity. Sampling methods provide asymptotically consistent estimates, but at a high computational cost. It is therefore beneficial to investigate the linearization approximation, which provides estimates at a much smaller computational cost. However, the accuracy of the linearization typically varies between different models, requiring evaluation before they can be applied.

% Previous work and how it relates to ours.
The problem of ML estimation in nonlinear SSMs is widely considered in literature. One method used by e.g.\ \cite{kokS:2014} approximates the log-likelihood in~\eqref{eq:ml} using an EKF. The ML estimation problem is subsequently solved using quasi-Newton optimization for which the gradients are approximated using finite differences. This results in a rapidly increasing computational complexity as the number of parameters grows. Other approaches are based on gradient ascent algorithms together with particle methods, see e.g.\ \citet{PoyiadjisDoucetSingh2011} and \citet{doucetJR:2013}. These approaches typically require the user to pre-define step sizes, which can be challenging in problems with a highly skewed log-likelihood. Lastly, gradient-free methods based on simultaneous perturbation stochastic approximation~\citep{EhrlichJasraKantas2015}, Gaussian processes optimization~\citep{DahlinLindsten2014} and expectation maximization~\citep{schonWN:2011,kokkalaSS:2014} can be used. However, these methods typically do not enjoy the quadratic convergence rate of Newton methods, see e.g.\ \cite{PoyiadjisDoucetSingh2011} for a discussion.

%%%%%%%%%%%%%%%%%
\section{Strategies for computing derivatives of the log-likelihood}
\label{sec:problemForm}
A Newton algorithm to solve~\eqref{eq:ml} iterates the update
\begin{equation}
\begin{split}
\label{eq:NewtonUpdate}
	\theta_{k+1} 
	&= 
	\theta_{k} 
	-
	\varepsilon_k 
	\Big[ \mathcal{H}(\theta_k) \Big]^{-1}
	\Big[ \mathcal{G}(\theta_k) \Big], 
	\\
	\mathcal{G}(\theta_k)
	&\triangleq 
	\tfrac{\partial}{\partial \theta} \log p_{\theta}(y_{1:N}) \big|_{\theta=\theta_k}, \\
	\mathcal{H}(\theta_k)
	&\triangleq 
	\mathbb{E}_{y_{1:N}} \Big[ \tfrac{\partial^2}{\partial^2 \theta} \log p_{\theta}(y_{1:N}) \big|_{\theta=\theta_k} \Big],
\end{split}
\end{equation}
where $\varepsilon_k$ denotes a step length determined for instance using a line search algorithm~\citep{nocedalW:2006} or using an update based on stochastic approximation~\citep{PoyiadjisDoucetSingh2011}. Here, we introduce the notation $\mathcal{G}(\theta_k)$ and $ \mathcal{H}(\theta_k)$ for the gradient and the Hessian of the log-likelihood, respectively. In Algorithm~\ref{alg:newton}, we present the full procedure for solving the \ac{ml} problem \eqref{eq:ml} using \eqref{eq:NewtonUpdate}.

To make use of Algorithm~\ref{alg:newton}, we require estimates of the gradient and the Hessian. We estimate the gradient via Fisher's identity,
\begin{align}
	\mathcal{G}(\theta) = \dint \tfrac{\partial}{\partial \theta} \log p_{\theta}(x_{1:N},y_{1:N}) p_{\theta}(x_{1:N} | y_{1:N}) \dd x_{1:N}.
	\label{eq:FishersIdentity}
\end{align}
Here, computation of the gradient of the log-likelihood amounts to a marginalization of the
gradient of the \textit{complete data log-likelihood} with respect to the states. As the states are
unknown, this marginalization cannot be carried out in closed form, which is the reason why the gradients cannot be computed exactly. The complete data likelihood for an \ac{ssm} is given by
\begin{align}
	p_{\theta}(x_{1:N},y_{1:N}) &= 
	p_{\theta}(x_1)
	\prod_{t=1}^{N-1} \fth (x_{t+1}|x_{t})
	\prod_{t=1}^N \gth (y_{t}|x_{t}),
	\label{eq:CompleteDataLikelihood}%
\end{align}%
where $p_{\theta}(x_1)$, $\fth (x_{t+1} | x_{t})$ and $\gth (y_t | x_t)$ are given by~\eqref{eq:ss}. In this paper, we assume that we can compute the gradient of the logarithm of this expression with respect to the parameter vector. By inserting \eqref{eq:CompleteDataLikelihood} into \eqref{eq:FishersIdentity}, we can make use of the Markov property of the model to obtain
\begin{align}
	&\mathcal{G}(\theta_k) 
	= 
	\sum_{t=1}^N 
	\underbrace{\dint \xi_{\theta_k}(x_{t+1:t}) p_{\theta_k}(x_{t+1:t}|y_{1:N}) \dd x_{t+1: t}}_{\triangleq \mathcal{G}_t(\theta_k)}, 
	\label{eq:FishersIdentityII} \\
	&\xi_{\theta_k}(x_{t+1:t}) \triangleq \tfrac{\partial}{\partial \theta} \log \fth (x_{t+1} | x_{t}) \big|_{\theta=\theta_k} + \tfrac{\partial}{\partial \theta} \log \gth (y_t | x_{t}) \big|_{\theta=\theta_k}, \nonumber%
\end{align}
where we interpret $\fth (x_1 | x_0)=p(x_1)$ for brevity. The remaining problem is to estimate the
two-step joint smoothing distribution $p_{\theta_k} (x_{t+1:t} | y_{1:N})$ and insert it into the
expression.

Furthermore, the Hessian can be approximated using the gradient estimates according to \citet{segalW:1989} and \citet{Meilijson1989}. The estimator is given by
\begin{align}
	\widehat{\mathcal{H}}(\theta_k) 
	&=
	\frac{1}{N}
	\Big[ \mathcal{G}(\theta_k) \Big]
	\Big[ \mathcal{G}(\theta_k) \Big]^{\top}
	-
	\sum_{t=1}^N
	\Big[ \mathcal{G}_t(\theta_k) \Big]\Big[ \mathcal{G}_t(\theta_k) \Big]^{\top}
	\label{eq:InfoEstimator}
\end{align}
which is a consistent estimator of the expected information matrix. That is, we have that $\widehat{\mathcal{H}}(\theta^{\star}) \rightarrow \mathcal{H}(\theta^{\star})$ as $N \rightarrow \infty$ for the true parameters $\theta^{\star}$ if the gradient estimate $\widehat{\mathcal{G}}(\theta)$ is consistent. The advantage of this estimator is that Hessian estimates are obtained as a by-product from the estimation of the gradient. 

Both estimators in \eqref{eq:FishersIdentityII} and \eqref{eq:InfoEstimator}, require estimates of the intractable two-step smoothing distribution. In the two subsequent sections, we discuss the details of how to make use of the linearization and sampling approximations to estimate this smoothing distribution. 

\begin{algorithm}[!t]
\caption{\textsf{Newton method for \ac{ml} parameter estimation}}
\textsc{Inputs:} Initial parameter $\theta_0$, maximum no.\ iterations $K$. \\
\textsc{Outputs:} \ac{ml} parameter estimate $\widehat{\theta}_{\text{ML}}$.
\algrule[.4pt]
\begin{enumerate}[label={\arabic*:}]
	\item Set $k=0$
	\item \textbf{while} \textit{exit condition is not satisfied} \textbf{do}
	\begin{enumerate}[label={\alph*:}]
		\item Run an algorithm to estimate the log-likelihood $\widehat{\ell}(\theta_k)$, its gradient $\widehat{\mathcal{G}}(\theta_k)$ and its Hessian $\widehat{\mathcal{H}}(\theta_k)$.
		\item Determine $\varepsilon_k$ using e.g.\ a line search algorithm or a stochastic schedule.
		\item Apply the Newton update \eqref{eq:NewtonUpdate} to obtain $\theta_{k+1}$.
		\item Set $k=k+1$.
	\end{enumerate}
    	\item[] \textbf{end while}
	\item Set $\widehat{\theta}_{\text{ML}} = \theta_{k}$.
\end{enumerate}
\label{alg:newton}
\end{algorithm}

\section{Linearization approximation}
\label{sec:lin}
To make use of linearization approximations to estimate the gradient, we treat \eqref{eq:FishersIdentityII} as an expected value of the form
\begin{align}
	\label{eq:generalGradient}
	\mathcal{G}(\theta_k) 
	= 
	\sum_{t=1}^N
	\mathbb{E}_{\theta_k} \Big[ 
	\xi_{\theta_k}(x_{t+1:t})
	\big| y_{1:N} \Big].
\end{align}
In Section~\ref{sec:lin:linG}, we first consider the linear case and recapitulate the calculation in
\cite{segalW:1989}. These results are extended to nonlinear \ac{ssm}s with Gaussian additive noise in Section~\ref{sec:lin:nlinG}.

\subsection{Linear models with additive Gaussian noise}
\label{sec:lin:linG}
The linear Gaussian \ac{ssm} is given by the following special case of~\eqref{eq:ss},
\begin{equation}
\begin{aligned}
\label{eq:ssLinGauss}
x_{t+1} &= F(\theta) x_t + v_t(\theta), \quad &v_t(\theta) \sim \mathcal{N}(0,Q(\theta)), \\
y_t &= G(\theta) x_t + e_t(\theta), \quad &e _t(\theta) \sim \mathcal{N}(0,R(\theta)). 
\end{aligned}%
\end{equation}%
For notational brevity, we assume that the initial state is distributed as $x_1 \sim \mathcal{N}(\mu_x, P_1)$ and is independent of the parameters $\theta$. For this model, we can express the \ac{ml} problem as
\begin{align}
\label{eq:mlLinGauss}
\widehat{\theta}^{\text{ML}} = \argmax_\theta \sum_{t = 1}^N \log \mathcal{N} \Big( y_t; \widehat{y}_{t | t-1}(\theta) , S_t (\theta) \Big),
\end{align}
where it is possible to compute the predictive likelihood $\widehat{y}_{t | t-1}(\theta)$ and its covariance $S_t (\theta)$ using a \ac{kf}. 

To solve~\eqref{eq:mlLinGauss} using Algorithm~\ref{alg:newton}, the gradient and Hessian of the log-likelihood need to be computed. Using~\eqref{eq:generalGradient}, we first note that the complete data log-likelihood can be expressed as
\begin{align}
&\log p_\theta (x_{1:N} , y_{1:N}) = -\tfrac{N-1}{2} \log \det Q - \tfrac{N}{2} \log \det R - \nonumber \\
&\quad \tfrac{1}{2}  \| x_1 - \mu_x \|_{P_1^{-1}}^2 - \tfrac{1}{2} \log \det P_1 - \nonumber \\
&\quad \tfrac{1}{2}\sum_{t=1}^{N-1}  \| x_{t+1} - F x_{t} \|_{Q^{-1}}^2 -
\tfrac{1}{2}\sum_{t=1}^N \| y_t - G x_{t} \|_{R^{-1}}^2,
\end{align}%
where the explicit dependency on $\theta$ has been omitted for notational simplicity. The gradient of the log-likelihood~\eqref{eq:generalGradient} can then be written as 
\begin{small}
\begin{align}
\label{eq:linearGradients}
\widehat{\mathcal{G}}(\theta) =& -\tfrac{N-1}{2} \Tr \left( Q^{-1} \tfrac{\partial Q}{\partial \theta} \right) -\tfrac{N}{2} \Tr \left( R^{-1} \tfrac{\partial R}{\partial \theta} \right)  - \nonumber \\
& \tfrac{1}{2} \Tr \left(\sum_{t=2}^N (P_{t | N} + \widehat{x}_{t | N} \widehat{x}_{t | N}^\Transp ) \tfrac{ \partial Q^{-1}}{\partial \theta} \right) - \nonumber \\
& \tfrac{1}{2} \Tr \left( \sum_{t=2}^N (P_{t-1 | N} + \widehat{x}_{t-1 | N} \widehat{x}_{t-1 | N}^\Transp) \tfrac{ \partial F^\Transp Q^{-1} F}{\partial \theta} \right) + \nonumber \\
& \Tr \left( \sum_{t=2}^N (P_{t-1,t | N} + \widehat{x}_{t-1 | N} \widehat{x}_{t | N}^\Transp) \tfrac{ \partial Q^{-1} F}{\partial \theta} \right) + \nonumber \\
&  \sum_{t=1}^N y_t^\Transp \tfrac{\partial R^{-1} G}{\partial \theta} \widehat{x}_{t | N} - \tfrac{1}{2}\sum_{t=1}^N y_t^\Transp \tfrac{\partial R^{-1}}{\partial \theta} y_t - \nonumber \\ 
& \tfrac{1}{2} \Tr \left( \sum_{t=1}^N (P_{t | N} + \widehat{x}_{t | N} \widehat{x}_{t | N}^\Transp) \tfrac{ \partial G^\Transp R^{-1} G}{\partial \theta} \right),
\end{align}%
\end{small}%
where
$\widehat{x}_{t | N}$ and $P_{t | N}$ denote the smoothed state
estimates and their covariances, respectively. The term $P_{t-1,t |
  N}$ denotes the covariance between the states $x_{t-1}$ and
$x_{t}$. These quantities can be computed using a Kalman smoother, see e.g.~\cite{rauchST:1965}.

\subsection{Nonlinear models with additive Gaussian noise}
\label{sec:lin:nlinG}
A slightly more general \ac{ssm} is given by
\begin{equation}
\begin{aligned}
\label{eq:ssNonlinGauss}
x_{t+1} &= \fth (x_t) + v_t(\theta), \quad &v_t(\theta) \sim \mathcal{N}(0,Q(\theta)), \\
y_t &= \gth (x_t) + e_t(\theta), \quad &e_t(\theta) \sim \mathcal{N}(0,R(\theta)). 
\end{aligned}
\end{equation}
For this model, the \ac{ml} problem assumes the same form as in Section~\ref{sec:lin:linG}, but the predictive likelihood $\widehat{y}_{t | t-1}(\theta)$ and its covariance $S_t (\theta)$ cannot be computed exactly. Instead, we replace them with estimates obtained from an \ac{ekf}. 

To compute the gradient and the Hessian of the log-likelihood, assume for a moment that we can compute the complete data log-likelihood exactly. The gradient~\eqref{eq:generalGradient} can then be computed exactly in two special cases:
\begin{enumerate}
\item When part of the \ac{ssm}~\eqref{eq:ssNonlinGauss} is linear and the parameters only enter in the linear part of the model. In this case, the gradients reduce to their respective linear parts in~\eqref{eq:linearGradients}.
\item When the expectation in~\eqref{eq:generalGradient} can be computed exactly even though the model is nonlinear, for example in the case of quadratic terms. 
\end{enumerate}

The assumption that the complete data log-likelihood can be computed exactly is clearly not true for nonlinear SSMs. However, good estimates of $\widehat{x}_{t | N}$ can be obtained by solving the optimization problem 
\begin{align}
\label{eq:mapStates}
\widehat{x}_{t | N} &= \argmax_{x_{1:N}} p_{\theta_k} (x_{1:N}, y_{1:N}). 
\end{align}
This is a nonlinear least-squares problem and can be solved efficiently using a standard
Gauss-Newton solver due to its inherent sparsity. The state covariances $P_{t | N}$ and $P_{t-1,t |
  N}$ can be approximated from the inverse of the approximate Hessian of the complete data
log-likelihood
\begin{equation}
\begin{split}
\label{eq:approxHess}
	P_{1:N,1:N | N}(\widehat{x}_{t | N}) &\approx \left( \Big[ \mathcal{J}(\widehat{x}_{t | N}) \Big] \Big[ \mathcal{J}(\widehat{x}_{t | N}) \Big]^\Transp \right)^{-1}, \\
	\mathcal{J}(\widehat{x}_{t | N}) &\triangleq \tfrac{\partial }{\partial x} p_{\theta_k}
                                           (x_{1:N},y_{1:N}) \Big|_{x_{1:N} = \widehat{x}_{1:N}},
\end{split}
\end{equation}
where $P_{1:N,1:N | N}$ represents the smoothed covariance matrix representing the covariance between all states. Since we are only interested in $P_{t | N}$ (which is short notation for $P_{t,t | N}$) and $P_{t-1,t | N}$, we are only interested in a few components of $P_{1:N,1:N | N}$. Hence, it is not necessary to form an explicit inverse, which would be intractable for large $N$. 

\begin{algorithm}[t]
\caption{\textsf{Computation of the log-likelihood and its derivatives using linearization approximation}}
\textsc{Inputs:} A parameter estimate $\theta_k$ \\
\textsc{Outputs:} An estimate of the log-likelihood $\widehat{\ell}(\theta_k)$ and its gradient $\widehat{\mathcal{G}}(\theta_k)$ and Hessian $\widehat{\mathcal{H}}(\theta_k)$.
\algrule[.4pt]
\begin{enumerate}[label={\arabic*:}]
	\item Run an \ac{ekf} using $\theta_k$ to obtain $\widehat{\ell}(\theta_k)$.
	\item Solve the smoothing problem~\eqref{eq:mapStates}.
	\begin{enumerate}[label={\alph*:}]
		\item Initialize $x^1_{1:N | N}$ using the state estimates from the \ac{ekf}.
		\item \textbf{while} \textit{exit condition is not satisfied} \textbf{do}
		\begin{enumerate}[label={\roman*:}]
			\item Compute the gradient $\mathcal{J}(x^i_{t | N})$ and approximate the Hessian of the smoothing problem using~\eqref{eq:approxHess}. 
			\item Use a Gauss-Newton update similar to~\eqref{eq:NewtonUpdate} to obtain $x^{i+1}_{t | N}$.
			\item Set $i=i+1$.
		\end{enumerate}
		\item[] \textbf{end while}
		\item Set $\widehat{x}_{1:N | N} = x^{i+1}_{1:N | N}$.
	\end{enumerate}
	\item Compute $P_{t,t | N}$ and $P_{t-1,t | N}$ using~\eqref{eq:approxHess}.
	\item Use~\eqref{eq:generalGradient} to estimate the gradient $\widehat{\mathcal{G}}(\theta_k)$ and use~\eqref{eq:InfoEstimator} to determine the Hessian $\widehat{\mathcal{H}}(\theta_k)$.
\end{enumerate}
\label{alg:ekf}
\end{algorithm}

\section{Sampling approximation}
\label{sec:particle}
An alternative approximation of the log-likelihood and the solution to the smoothing problem uses sampling methods. The approximation is based on particle filtering and smoothing and can be applied to more general nonlinear \ac{ssm}s than the special cases introduced in Section~\ref{sec:lin}. 

Particle methods are a specific instance of sequential Monte Carlo (SMC; \citealp{DoucetJohansen2011}) methods, when this family of algorithms is applied on general \ac{ssm}s. The \ac{pf} is a combination of sequential importance sampling and resampling applied to estimate the states of an SSM. For example, the two-step smoothing distribution required in \eqref{eq:FishersIdentityII} can be estimated by
\begin{align}
	\widehat{p}_\theta (\dn x_{t+1:t} | y_{1:N})
	\triangleq
	\sum_{i=1}^M
	w_{N}^{(i)}
	\delta_{x_{t+1:t}^{(i)}}
	(\dn x_{t+1:t}),
	\label{eq:PFtwoStepEstimator}
\end{align}
where $w_t^{(i)}$ and $x_t^{(i)}$ denote the weights and locations of the particles obtained by the \ac{pf}. The particle system $\Gamma_M = \big\{ x_{1:N}^{(i)},w_{1:N}^{(i)} \big\}_{i=1}^M$ required to compute \eqref{eq:PFtwoStepEstimator} is obtained by an iterative algorithm with three steps: resampling, propagation and weighting. The most commonly used \ac{pf} is the \textit{bootstrap particle filter} (bPF), which is summarized in Step 1 of Algorithm~\ref{alg:bpf}.

\begin{algorithm}[t]
\caption{\textsf{Computation of the log-likelihood and its derivatives using sampling approximation}}
\textsc{Inputs:} Parameter estimate $\theta_k$ and no.\ particles $M$. \\
\textsc{Outputs:} Estimate of gradient $\widehat{\mathcal{G}}(\theta_k)$ and Hessian $\widehat{\mathcal{H}}(\theta_k)$.
\algrule[.4pt]
\begin{enumerate}[label={\arabic*:}]
	\item Run a bPF using $\theta_k$ to estimate the particle system. \\
	\textit{(all operations are carried out over $i,j = 1, \ldots, M$)}
	\begin{enumerate}[label={\alph*:}]
		\item Sample $x^{(i)}_1 \sim p_{\theta_k}(x_1)$ and compute $w_1^{(i)}$ by Step iii.
		\item \textbf{for} $t=2$ to $N$ \textbf{do} 
		\begin{enumerate}[label={\roman*:}]
			\item \textsf{Resampling}: sample a new ancestor $a^{(i)}_t$ from a multinomial distribution with
			$\mathbb{P} \big( a^{(i)}_t = j \big) = w^{(j)}_{t-1}$.
			\item \textsf{Propagation}: Sample
			$x_t^{(i)} \sim f_{\theta_k} \big( x_t^{(i)} \big| x_{t-1}^{a^{(i)}_t} \big)$.
			\item \label{alg:step:weights} \textsf{Weighting}: Calculate
			$\bar{w}^{(i)}_t = g_{\theta_k} \big( y_t|x_t^{(i)} \big)$, which by normalisation (over $i$) gives $w^{(i)}_t$.
		\end{enumerate}
	\item[] \textbf{endfor}
	\end{enumerate}
	\item Run a smoothing algorithm, e.g.\ FL or FFBSi.
	\item Use~\eqref{eq:FLScoreEst} or~\eqref{eq:FFBSiScoreEst} together with \eqref{eq:InfoEstimator} to estimate the gradient $\widehat{\mathcal{G}}(\theta_k)$ and the Hessian $\widehat{\mathcal{H}}(\theta_k)$.
\end{enumerate}
\label{alg:bpf}
\end{algorithm}

The estimator in \eqref{eq:PFtwoStepEstimator} can be directly inserted into
\eqref{eq:FishersIdentityII} to estimate the gradient of the log-likelihood. However, this typically
results in estimates suffering from high variance due to the problem of \textit{path
  degeneracy}. This is a result of the successive resamplings of the particle system (Step~i in
Algorithm~\ref{alg:bpf}) that collapse the particle trajectories. Hence, all particles share a
common ancestor, i.e.\ effectively we have $M \approx 1$.

% Discuss different particle smoothers
Alternative methods instead rely on \emph{particle smoothing} to estimate the two-step smoothing distribution, which results in estimators with better accuracy but with an increase in the computational complexity. In this paper, we consider two different particle smoothers: the fixed-lag (FL; \citealp{KitagawaSato2001}) smoother and the forward filter backward simulator (FFBSi). There are numerous other alternatives and we refer to \citet{LindstenSchon2013} for a survey of the current state-of-the-art.% in particle smoothing. 
\acused{FFBSi} \acused{FL}

\subsection{Fixed-lag smoothing}
\label{sec:partMethods:FL}
% Introduce the underlying assumption of the FL smoother and give the emperical 2-step filtering distribution
The \ac{FL} smoother relies upon the assumption that the \ac{ssm} is mixing fast and forgets about its past within a few time steps. More specifically, this means that we can approximate the two-step joint smoothing distribution by
\begin{align}
	\widehat{p}_{\theta_k} (\dn x_{t+1:t}|y_{1:N}) \approx \widehat{p}_{\theta_k} (\dn x_{t+1:t}|y_{1:\kappa_t}),
\end{align}
where $\kappa_t = \max(N,t+1+\Delta)$ for some fixed-lag $0 < \Delta \leq N$. The resulting estimator has the form
\begin{align}
	\widehat{p}_{\theta_k} (\dn x_{t+1:t}| y_{1:N}) 
	\triangleq 
	\sum_{i=1}^M 
	w_{\kappa_t}^{(i)} 
	\delta_{\tilde{x}_{\kappa_t,t+1:t}^{(i)}}
	(\dn x_{t+1:t}),
	\label{eq:FL2step}
\end{align}
where $\tilde{x}_{\kappa_t,t}^{(i)}$ denotes the ancestor at time~$t$ of particle $x_{\kappa_t}^{(i)}$. The gradient can subsequently be estimated by inserting \eqref{eq:FL2step} into \eqref{eq:FishersIdentityII} to obtain
\begin{align}
    \widehat{\mathcal{G}}(\theta_k)
	=
	\sum_{t=1}^N
	\sum_{i=1}^M	
	w_{\kappa_t}^{(i)}
	\xi_{\theta_k} \Big( \tilde{x}_{\kappa_t,t+1:t}^{(i)} \Big).
	\label{eq:FLScoreEst}	
\end{align}
% Discuss the statistical properties of the FL smoother
The implementation of this expression is straightforward and is summarized in Algorithm~\ref{alg:bpf}. See \cite{Dahlin2014} for the complete derivation and algorithm of the \ac{FL} smoother. The \ac{FL} smoother retains the computational complexity of the \ac{pf}, i.e.\ $\mathcal{O}(NM)$. Furthermore, it enjoys improved statistical properties under some general mixing assumptions of the SSM. A drawback with the \ac{FL} smoother is a persisting bias that does not vanish in the limit when $M \rightarrow \infty$. 

\subsection{Forward filter backward simulator}
\label{sec:partMethods:FFBSi}
% Introduce the computational cost and statistical properties of FFBSi
The second smoother that we consider is the \ac{FFBSi} algorithm, which enjoys even better statistical properties than the \ac{FL} smoother but with a computational complexity proportional to $\mathcal{O}(NM \bar{M})$, where $\bar{M}$ denotes the number of particles in the backward sweep. The estimates obtained with the \ac{FFBSi} smoother are asymptotically consistent and therefore unbiased in the limit when $M,\bar{M} \rightarrow \infty$.

% Give the resulting estimator for the gradient
The \ac{FFBSi} algorithm can be seen as the particle approximation of the \ac{rts} smoother. The gradient can be estimated using a backward sweep after a forward pass using bPF in Algorithm~\ref{alg:bpf}. In the backward sweep, we draw $\bar{M}$ trajectories $\big\{ \tilde{x}_{1:N}^{(j)} \big\}_{j=1}^{\bar{M}}$ by rejection sampling from the backward kernel given by
\begin{align}
    \widehat{B}_t( \dn x_t | x_{t+1} )
    \triangleq
    \sum_{i=1}^M
    \frac{ w_t^{(i)} f_{\theta_k}( x_{t+1} | x_t^{(i)} ) }
    { \sum_{l=1}^M w_t^{(l)} f_{\theta_k}( x_{t+1} | x_t^{(l)} ) }
    \delta_{x_t^{(i)}} ( \dn x_t ),
\end{align}
where $\Gamma_M$ generated by the forward pass. The gradient can subsequently be estimated as
\begin{align}
    \widehat{\mathcal{G}}(\theta_k)
	=
	\frac{1}{\bar{M}}
	\sum_{t=1}^N
	\sum_{i=1}^{\bar{M}}
	\xi_{\theta_k} \Big( \tilde{x}_{t+1:t}^{(i)} \Big).
	\label{eq:FFBSiScoreEst}	
\end{align}

To decrease the computational time, we make use of the early stopping rule discussed by \cite{TaghaviLindstenSvenssonSchon2013}. Here, we sample the first $M_{\text{limit}}$ backward trajectories using rejection sampling and the remaining using standard FFBSi. This results in a computationally efficient and accurate smoothing algorithm. See Algorithm~6 in \cite{LindstenSchon2013} for the complete algorithm and a discussion of its statistical properties.

\section{Simulation results}
\label{sec:results}
We evaluate the proposed methods by considering two different \ac{ssm}s. For both models, we compare the estimates obtained using four different methods: 
\begin{enumerate}
\item \textbf{ALG2: } The Newton-based optimization in Algorithm~\ref{alg:newton} in combination with estimates of the log-likelihood and its derivatives from Algorithm~\ref{alg:ekf}.
\item \textbf{ALG3FL: } The Newton-based optimization in Algorithm~\ref{alg:newton} in combination
  with estimates of the log-likelihood and its derivatives from Algorithm~\ref{alg:bpf} using an
  \ac{FL} smoother, with $\Delta=12$ and $M=2 \thinspace 000$. We use the sequence of step sizes given by $\varepsilon_k = k^{-2/3}$ as suggested by \cite{PoyiadjisDoucetSingh2011}.
\item \textbf{ALG3FFBSi: } Similar to ALG3FL but instead of an \ac{FL} smoother we use an \ac{FFBSi} smoother, with $M = 2 \thinspace 000$, $\bar{M}=100$ and $M_{\text{limit}} = 10$.
\item \textbf{NUM: } A quasi-Newton algorithm, which computes the log-likelihood using an \ac{ekf}
  as in Section~\ref{sec:lin:nlinG}, but the gradients are found using finite-differences of the approximate 
  log-likelihood instead.
\end{enumerate} 

The latter method is included in the comparison because it is commonly used in ML parameter estimation approaches where the log-likelihood is approximated using linearization, see e.g. \citet{kokS:2014} for a concrete example. It is computationally expensive for large dimensions of the parameter vector, but is cheap for the low dimensional parameter vectors we consider here. Source code for the simulated examples is available via the first author's homepage.\footnote{http://users.isy.liu.se/en/rt/manko/}

\subsection{Parameters in the linear part of the \ac{ssm}}
\label{sec:results:model1}
\begin{figure}[p]
\centering
\includegraphics[width=0.98\columnwidth]{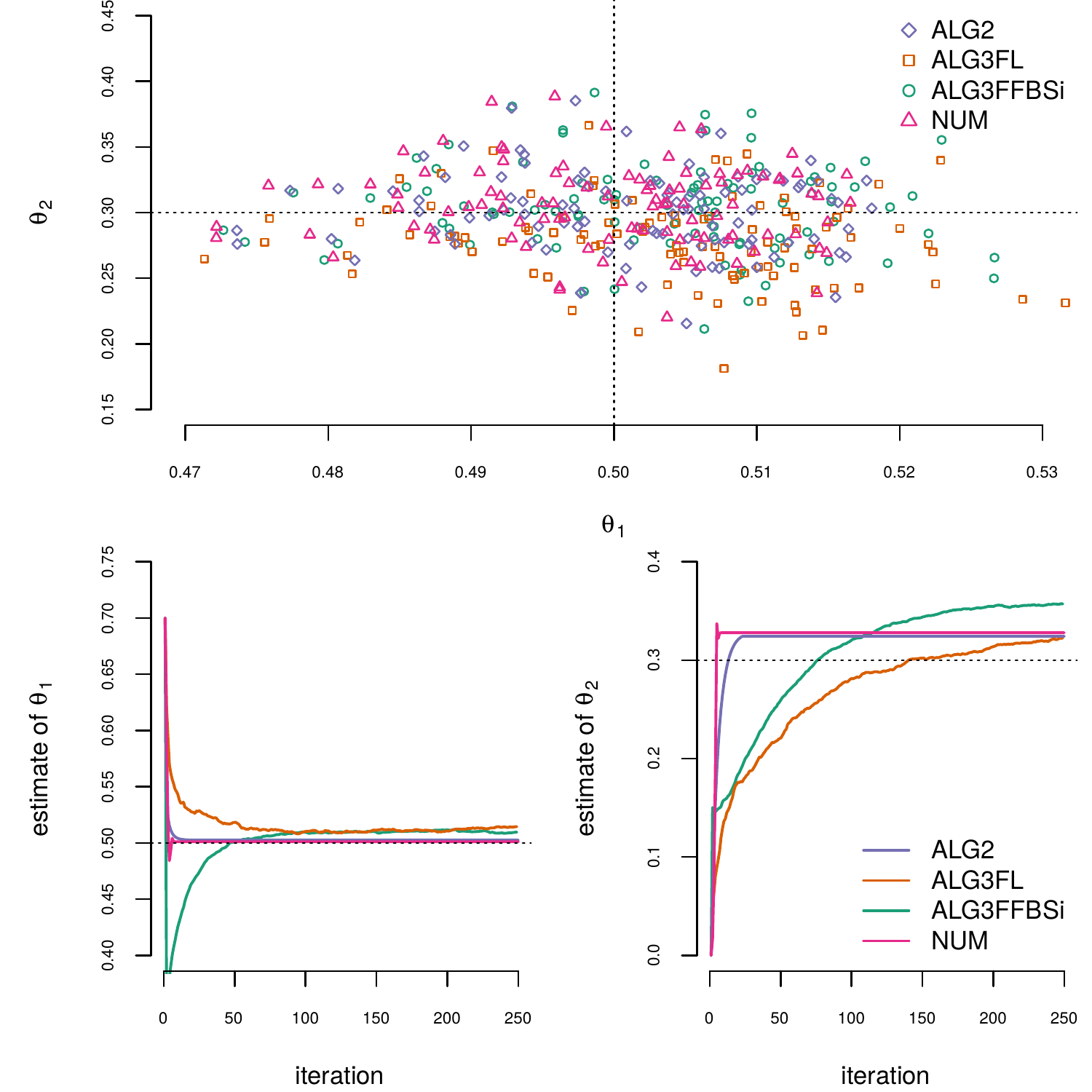}
\caption{Comparison of the different methods for the \ac{ssm}~\eqref{eq:model1}. Upper: The estimates of $\theta_1$ and $\theta_2$ using $100$ data sets. Lower: trace plots of the optimization methods for one of the data sets for $\theta_1$ (left) and $\theta_2$ (right).}
\label{fig:results-model1}
\end{figure}

The first model that we consider is given by
\begin{equation}
\begin{aligned}
\label{eq:model1}%
x_{t+1} &= \arctan x_t + v_t, \quad &v_t &\sim \mathcal{N}(0,1),\\
y_t &= \theta_1 x_t + \theta_2 + e_t, \quad &e_t &\sim \mathcal{N}(0,0.1^2).
\end{aligned}%
\end{equation}%
The unknown parameter vector is $\theta = \left\{ \theta_1, \theta_2 \right\}$. The model~\eqref{eq:model1} has nonlinear dynamics, but the measurement equation is linear. This corresponds to a model of type 1 as discussed in Section~\ref{sec:lin:nlinG}. As the parameters are located in the linear measurement equation, the expressions for the gradient~\eqref{eq:generalGradient} and Hessian~\eqref{eq:InfoEstimator} in Algorithm~\ref{alg:ekf} are equal to their linear counterparts in~\eqref{eq:linearGradients}.

We simulate $100$ data sets each consisting of $N=1 \thinspace 000$ samples and true parameters $\theta^{\star}=\{0.5,0.3\}$ to compare the accuracy of the proposed methods. All methods are initialized in $\theta_0=\{0.7,0.0\}$. The parameter estimates from the four methods are presented in the upper plot in Fig.~\ref{fig:results-model1}. The bias and mean square errors (MSEs) as compared to $\theta^\star$ are also represented in Table~\ref{tbl:results-lgssm}. Note that in the model~\eqref{eq:model1}, positive and negative $\theta_{k,1}$ are equally likely. Hence, without loss of generality we mirror all negative solutions to the positive plane. 

To illustrate the convergence of the different methods, the parameter estimates as a function of the iterations in the Newton method are shown in the lower plot in Fig.~\ref{fig:results-model1}. As can be seen, all four methods converge, but ALG2 and NUM require far less iterations than the ALG3FL and ALG3FFBSi.

For the \ac{ssm}~\eqref{eq:model1} it can be concluded that ALG2 outperforms the other methods both in terms of the bias and the MSE. The reason is that, as argued in Section~\ref{sec:lin:nlinG}, accurate gradient and Hessian estimates can be obtained for this model because of its structure. Note that ALG2 and NUM not only require the least computational time per iteration but also require far less iterations to converge than ALF3FL and ALG3FFBSi, making them by far the most computationally efficient algorithms. 

\begin{figure}[p]
\centering
\includegraphics[width=0.98\columnwidth]{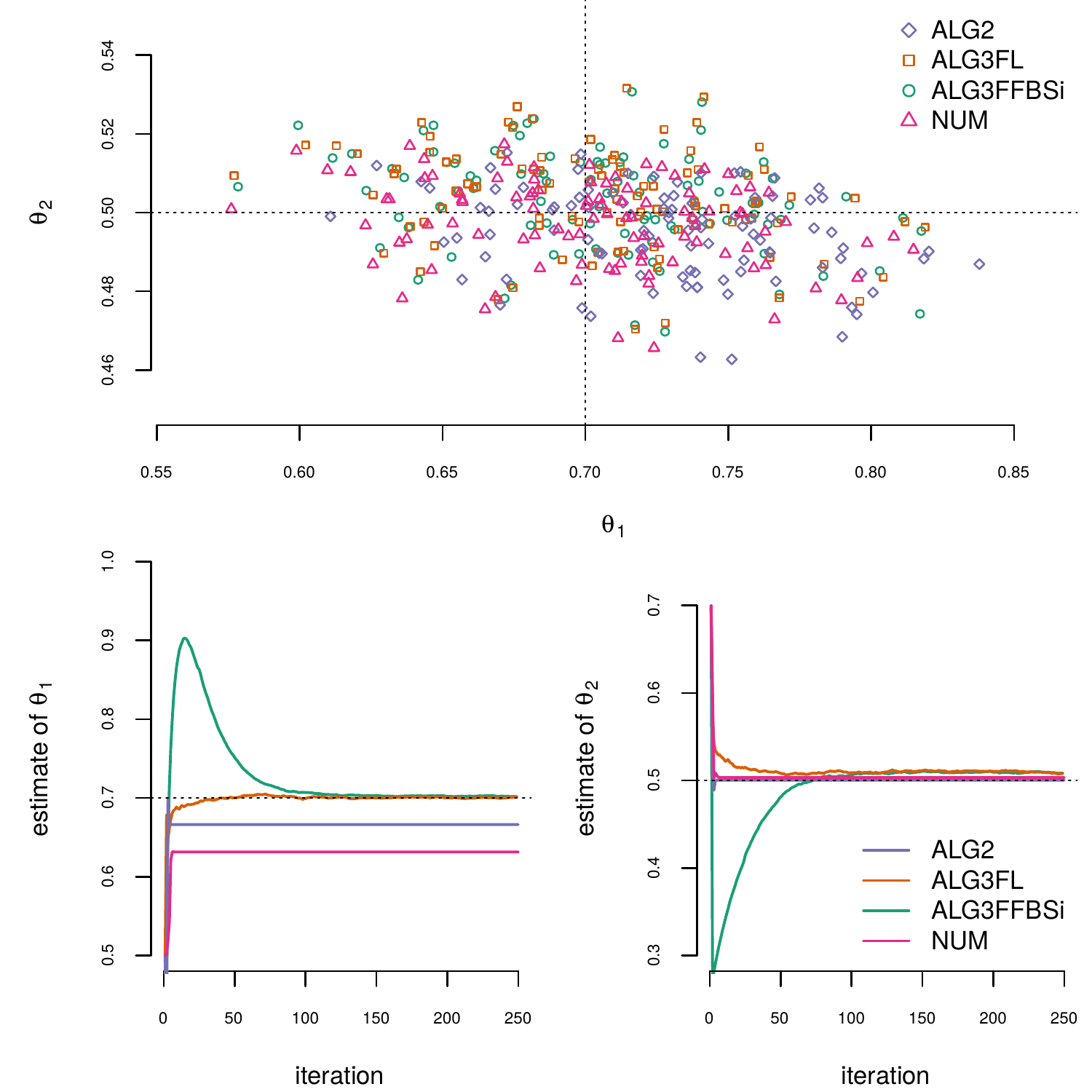}
\caption{Comparison of the different methods for the \ac{ssm}~\eqref{eq:model2}. Upper: The estimates of $\theta_1$ and $\theta_2$ using $100$ data sets. Lower: trace plots of the optimization methods for one of the data sets for $\theta_1$ (left) and $\theta_2$ (right).}
\label{fig:results-model2}
\end{figure}

\begin{table}[t]
\caption{The bias, MSE as compared to $\theta^\star$ and the computational time averaged over $100$ data sets of $N = 1 \thinspace 000$ for the \ac{ssm}s~\eqref{eq:model1} (model~1) and~\eqref{eq:model2} (model~2). The boldface font indicates the smallest MSE and bias.}
\begin{center}
\begin{tabular}{llccccc}
\toprule
&Alg. & \multicolumn{2}{c}{Bias ($\cdot 10^{-4}$)} & \multicolumn{2}{c}{MSE ($\cdot 10^{-4}$)} & Time\\
\cmidrule(r){3-4}
\cmidrule(r){5-6}
\cmidrule(r){7-7}
&& $\theta_1$ & $\theta_2$ & $\theta_1$ & $\theta_2$& (sec/iter) \\
\midrule
\parbox[t]{2mm}{\multirow{4}{*}{\rotatebox[origin=c]{90}{Model 1}}}
&ALG2          &  10  &  \textbf{10}  &\textbf{1}  & \textbf{10} & 0.81 \\
&ALG3FL        &  38  & -214 & 2  & 16 & 4 \\
&ALG3FFBSi     &  31  &  53  & \textbf{1}  & 11 & 19 \\
&NUM           & \textbf{-4}   &  48  & \textbf{1}  & \textbf{10} & 0.19 \\
\midrule
\parbox[t]{2mm}{\multirow{4}{*}{\rotatebox[origin=c]{90}{Model 2}}}
&ALG2          &  284 & -55 &  28 & 2 & 2.73 \\
&ALG3FL        &  53  &   35 & 24  & 2 & 5 \\
&ALG3FFBSi     &  55  &   \textbf{31} & 24  & 2 & 22 \\
&NUM           &  \textbf{31}  &  -27 & \textbf{23}  & \textbf{1} & 0.19 \\
\bottomrule
\end{tabular}
\end{center}
\label{tbl:results-lgssm}
\end{table}

\subsection{Parameters in the nonlinear part of the \ac{ssm}}
\label{sec:results:model2}
The second model that we consider is given by
\begin{equation}
\begin{aligned}
\label{eq:model2}%
x_{t+1} &= \theta_1 \arctan x_t + v_t, \quad &v_t &\sim \mathcal{N}(0,1)\\
y_t &= \theta_2 x_t + e_t, \quad &e_t &\sim \mathcal{N}(0,0.1^2),
\end{aligned}%
\end{equation}%
\noindent where $\theta_1$ scales the current state in the nonlinear state dynamics. We apply the same evaluation procedure as for the previous model but with $\theta^{\star}=\{0.7,0.5\}$ and $\theta_0=\{0.5,0.7\}$.

Note that for this case, the gradients for ALG2 can not be computed analytically. Denoting $\theta_{k,1}$ the estimate of $\theta_1$ at the $k^\text{th}$ iteration, we approximate the gradient with respect to $\theta_{k,1}$ by
\begin{align}
\widehat{\mathcal{G}}(\theta_{k,1}) =& \sum_{t=2}^N \mathbb{E}_{\theta_{k,1}} \Big[ \tfrac{\partial}{\partial \theta_{k,1}} \log p_{\theta_{k,1}} (x_t | x_{t-1}) | y_{1:N} \Big] \nonumber \\
=& \sum_{t=2}^N \mathbb{E}_{\theta_{k,1}} \Big[ - \tfrac{1}{2} \left( x_{t} - \theta_{k,1} \arctan x_{t-1} \right)^2 | y_{1:N} \Big]  \nonumber \\
\approx& \sum_{t=2}^N - \tfrac{1}{2} \left( \widehat{x}_{t | N} - \theta_{k,1} \arctan \widehat{x}_{t-1 | N} \right)^2.
\label{eq:model2approx}
\end{align}
Due to the necessity of the approximation~\eqref{eq:model2approx}, we expect ALG2 to perform worse for the SSM~\eqref{eq:model2} as compared to~\eqref{eq:model1}. This is confirmed by the results shown in Fig.~\ref{fig:results-model2} and summarized in Table~\ref{tbl:results-lgssm}. For this model, NUM outperforms the other methods both in bias, MSE and computational time. 

\section{Conclusions}
We have considered the problem of \ac{ml} parameter estimation in nonlinear \ac{ssm}s using Newton methods. We determine the gradient and Hessian of the log-likelihood using Fisher's identity in combination with an algorithm to obtain smoothed state estimates. We have applied this method to simulated data from two nonlinear SSMs. The first SSM has nonlinear dynamics, but the measurement equation is linear. The parameters only enter the measurement equation. Because of this, accurate parameter estimates are obtained using linearization approximations of the log-likelihood and its gradient and Hessian. In the second SSM, however, the parameters enter in both the linear and in the nonlinear part of the SSM. Because of this, the linearization approximations of the gradient and Hessian are of poor quality, leading to worse parameter estimates. For this SSM, the methods based on sampling approximations perform considerably better. However, a computationally cheap quasi-Newton algorithm which computes the log-likelihood using an EKF and finds the gradients using finite-differences outperforms the other methods on this example. Note that this algorithm is only computationally cheap for small dimensions of the parameter vector. Furthermore, it highly depends on the quality of the EKF state estimates. 

In future work, we would like to study the quality of the estimates for a wider range of nonlinear models. It would
also be interesting to compare the method introduced in this work to a solution based on expectation maximization. 
Furthermore, solving optimization problems by making use of the noisy estimates of the gradient and the Hessian 
that are provided by the particle smoother is an interesting and probably fruitful direction for future work.

%%%%%%%%%%%%%%%%%%%%%%%%%%%%%%%%%%%%%%%%%%%%%%%%%%%%%%%%%%%%%%%%%%%%%%%%%%%%%%%%%%%%%%%%%%%%%%%%%%%%%%%
%%%%%%%%%%%%%%%%%%%%%%%%%%%%%%%%%%%%%%%%%%%%%%%%%%%%%%%%%%%%%%%%%%%%%%%%%%%%%%%%%%%%%%%%%%%%%%%%%%%%%%%
%%%%%%%%%%%%%%%%%%%%%%%%%%%%%%%%%%%%%%%%%%%%%%%%%%%%%%%%%%%%%%%%%%%%%%%%%%%%%%%%%%%%%%%%%%%%%%%%%%%%%%%
\newpage
\bibliographystyle{plainnat}
\bibliography{references}

\end{document}